\documentclass[proceedings,opts]{JHEP}
\usepackage{epsfig} 

\def\Br{{\cal B}}
\def\bbbar{{\rm B}\bar{\rm B}}
\def\BZ{{\rm B}^0}
\def\BZB{\bar{\rm B}^0}
\def\BP{{\rm B}^+}
\def\BM{{\rm B}^-}
\def\KS{{\rm K}_{\rm s}^0}
\def\KZ{{\rm K}^0}
\def\PZ{\pi^0}
\def\KP{{\rm K}^+}
\def\PP{\pi^+}

\def\PM{\pi^-}
\def\ra{\rightarrow} 

\def\EB{E_{\rm B}}
\def\EBEAM{E_{\rm beam}}
\def\MB{M_{\rm B}}

\def\CLEOBSGAMMA{(3.15\pm{0.35}\pm{0.32}\pm{0.26})\times 10^{-4}}
\def\ALEPHBSGAMMA{(3.11\pm{0.80}\pm{0.72})\times 10^{-4}}

\def\KPPM{{\rm K}^+\pi^-}
\def\PPPM{\pi^+\pi^-}
\def\KPKM{{\rm K}^+{\rm K}^-}

\def\YBZTOKPPM{80.2{}^{+11.8}_{-11.0}} 
\def\YBZTOPPPM{20.0{}^{+7.6}_{-6.5}} 

\def\UBRBZTOKPPM{1.88\ {}^{+0.28}_{-0.26}\pm 0.13} 
\def\UBRBZTOPPPM{0.47\ {}^{+0.18}_{-0.15}\pm 0.06} 
\def\UBRBZTOKPKM{0.2} 

\def\BRBZTOKPPM{(\UBRBZTOKPPM)\times 10^{-5}} 
\def\BRBZTOPPPM{(\UBRBZTOPPPM)\times 10^{-5}} 
\def\BRBZTOKPKM{\UBRBZTOKPKM\times 10^{-5}} 

\def\KZPP{{\rm K}^0\pi^+}   
\def\KZKP{{\rm K}^0{\rm K}^+} 

\def\UBRBPTOKZPP{1.82{}^{+0.46}_{-0.40}\pm 0.16} 
\def\UBRBPTOKZKP{0.51} 

\def\BRBPTOKZPP{(\UBRBPTOKZPP)\times 10^{-5}} 
\def\BRBPTOKZKP{\UBRBPTOKZKP\times 10^{-5}} 

\def\KPPZ{{\rm K}^+\pi^0}   
\def\PPPZ{\pi^+\pi^0} 

\def\UBRBPTOKPPZ{1.21{}^{+0.30}_{-0.28}{}^{+0.21}_{-0.14}} 
\def\UBRBPTOPPPZ{1.2} 
\def\BRBPTOKPPZ{(\UBRBPTOKPPZ)\times 10^{-5}} 
\def\BRBPTOPPPZ{\UBRBPTOPPPZ\times 10^{-5}} 

\def\RZPP{\rho^0\pi^+}
\def\RZKP{\rho^0{\rm K}^+}
\def\YBPTORZPP{26.1{}^{+9.1}_{-8.0}}
\def\YBPTORZKP{14.8{}^{+8.8}_{-7.7}}
\def\UBRBPTORZPP{1.5\pm 0.5\pm 0.4} 
\def\UBRBPTORZKP{2.2}
\def\BRBPTORZPP{(\UBRBPTORZPP)\times 10^{-5}} 
\def\BRBPTORZKP{\UBRBPTORZKP\times 10^{-5}}

\def\RMPP{\rho^\pm\pi^\mp}
\def\RMKP{\rho^\pm{\rm K}^\mp}

\def\UBRBZTORMPP{3.5\ {}^{+1.1}_{-1.0}\pm 0.5} 
\def\UBRBZTORMKP{2.5}
\def\BRBZTORMPP{(\UBRBZTORMPP)\times 10^{-5}} 
\def\BRBZTORMKP{\UBRBZTORMKP\times 10^{-5}}

\def\KSTPPM{{\rm K}^{*\pm}\pi^\mp}
\def\KSTPKM{{\rm K}^{*\pm}{\rm K}^\mp}

\def\UBRBZTOKSTPPM{2.2\ {}^{+0.8}_{-0.6}\ {}^{+0.4}_{-0.5}} 
\def\UBRBZTOKSTPKM{0.6}

\def\OMGPP{\omega\pi^+}
\def\OMGKP{\omega{\rm K}^+}
\def\YBPTOOMGPP{28.5{}^{+8.2}_{-7.3}}
\def\YBPTOOMGKP{7.9{}^{+6.0}_{-4.7}}
\def\UBRBPTOOMGPP{1.1\pm0.3\pm0.1}
\def\UBRBPTOOMGKP{0.8}
\def\BRBPTOOMGPP{(\UBRBPTOOMGPP)\times 10^{-5}}
\def\BRBPTOOMGKP{\UBRBPTOOMGKP\times 10^{-5}}

\def\ETAPKZ{\eta'\KZ}
\def\ETAPKP{\eta'{\rm K}^+}
\def\ETAPPP{\eta'\pi^+}
\def\UBRBPTOETAPKP{8.0{}^{+1.0}_{-0.9}\pm0.8}
\def\UBRBZTOETAPKZ{8.8{}^{+1.8}_{-1.6}\pm0.9}
\def\UBRBPTOETAPPP{1.1}
\def\BRBPTOETAPKP{(\UBRBPTOETAPKP)\times 10^{-5}}
\def\BRBZTOETAPKZ{(\UBRBZTOETAPKZ)\times 10^{-5}}
\def\BRBPTOETAPPP{\UBRBPTOETAPPP\times 10^{-5}}

\def\UBRBPTOETAPKSTP{8.7}
\def\UBRBZTOETAPKSTZ{2.0}

\def\UBRBPTOETAKSTP{2.73{}^{+0.96}_{-0.82}\pm 0.50}
\def\UBRBZTOETAKSTZ{1.38{}^{+0.55}_{-0.44}\pm 0.17}

\def\UBRBPTOETAKP{0.71}
\def\UBRBZTOETAKZ{0.95}

\newcommand\nima[3]   { 
                {{\it Nucl.\ Inst.\ Meth.\ {\bf A}\ }{\bf #1} (#2) #3}}

\conference{Heavy Flavours 8, Southampton, UK, 1999}
\title{Recent Results on Rare $b$ Decays}
\author{David E. Jaffe\\ 
	University of California San Diego\\ 
	E-mail: \email{djaffe@lns.cornell.edu}}
\abstract{Recent experimental results on 
	charmless rare $b$ decays are reviewed.}
\begin{document}

\section{Introduction} 
Recent experimental results on rare $b$ decays are dominated by the
CLEO experiment with notable contributions from ALEPH, CDF and D0.
I will discuss only results on charmless hadronic B meson
decays ($B\to PP,PV$ where $P=\pi$, ${\rm K}$, $\eta$, $\eta'$ 
and $V = \rho$, ${\rm K}^*$, $\omega$), radiative $b$ decays
and purely leptonic B meson decays . 
I will omit any discussion of such rare $b$ decays
as ${\rm B}^+\to {\rm D}^0{\rm K}^+$,
${\rm B}^+\to {\rm D}^{*+}{\rm D}^{*-}$,  ${\rm B}^+\to \ell^+\nu$,
and $b\to s\nu\bar{\nu}$.

The search for rare $b$ decays lead to the first observation
of penguin decays in exclusive radiative B meson decays~\cite{ref:cleo.BtoKstG}
and revealed the unexpectedly large role of penguin contributions in
charmless hadronic B decays. The interference between the tree and
penguin contributions means that 
B decay rates to charmless hadronic final states are sensitive
to $\gamma$, the phase of ${\rm V}_{ub}$~\cite{ref:pdg}.
Several authors have proposed methods to bound $\gamma$ using ratios
of ${\rm B}\to {\rm K}\pi$ decay rates~\cite{ref:fm,ref:nr} with relatively
little model dependence. Both methods use 
${\rm B}\to {\rm K}\pi$ branching fractions to set bounds
on $\gamma$. Using the preliminary results contained in this paper,
neither the Fleischer-Mannel~\cite{ref:fm} bound
\begin{eqnarray*}
\Gamma({\rm B}^0\to {\rm K}^-\pi^+)/\Gamma({\rm B}^+\to {\rm K}^0\pi^+)
&\ge& \sin^2\gamma \\
1.11\pm0.35&=& \sin^2\gamma 
\end{eqnarray*}
nor the
Neubert-Rosner~\cite{ref:nr} bound
\begin{eqnarray*}
2R_* \equiv \Br(\BP\to\KZPP)/\Br(\BP\to\KPPZ) \\
(1-\sqrt{R_*})/\bar\epsilon_{3/2} \le |\delta_{\rm EW} - \cos\gamma| \\
0.55\pm0.74 \le |(0.64\pm0.15) - \cos\gamma|
\end{eqnarray*}
can set a meaningful
limit on $\gamma$~\cite{ref:explain}. 
Another method seeks to provide information on
$\gamma$ by sacrificing model-independence for comprehensive use of
existing measured branching fractions~\cite{ref:hsw}.

 In addition to probing $\arg({\rm V}_{ub}^*)$, the measurement of the proper
time dependence of rare B decays to $\pi\pi$~\cite{ref:pipi.alpha}
and $\pi\pi\pi$~\cite{ref:snyder.quinn} can yield a measurement
of $\alpha$~\cite{ref:pdg}. Finally, the search for direct $CP$ violation
in charmless hadronic and radiative B decays could provide evidence
of non-Standard Model (SM) physics.

 Table~\ref{tab:nb} lists the number of $b$ hadrons accumulated by different
experiments. 
The majority of results on charmless hadronic and radiative
$b$ decays come from the CLEO experiment that operates just above
the $\bbbar$ threshold at the $\Upsilon(4{\rm S})$ resonance. 
At $\sqrt{s} \approx 10.6 \ {\rm GeV}$,
the main source of background is continuum ${\rm e}^+{\rm e}^-\to q\bar{q}$
($q = ucsd$) with a cross-section of $\sim 3 \ {\rm nb}$ or about three
times $\sigma(\bbbar)$. A number of features allow the suppression or
successful treatment of the continuum by CLEO. Near threshold, the decay
products of the $\bbbar$ pair are isotropically distributed in contrast to
the back-to-back or ``jetty'' nature of continuum events. CLEO 
exploits the fact that the B meson energy $\EB$ is the beam energy $\EBEAM$ 
to form the beam-constrained mass ${\MB} \equiv \sqrt{{\EBEAM}^2 - {\vec {p}}_{\rm B}^{\ 2}}$
which is essentially a measure of the momentum balance of the B candidate
and has a resolution $\sigma({\MB}) \approx 2.5 \ {\rm MeV}$ dominated by
the beam energy spread. CLEO also takes approximately one third of its data
about 60 MeV below the $\Upsilon(4{\rm S})$ resonance, effectively turning
off the production of B mesons and permitting direct evaluation of continuum
background processes.
\TABLE[hbpt]{\begin{tabular}{lccc}
           & $\sqrt{s} $   & Production & Data  \\
Experiment & $({\rm GeV})$ & rate       &  sample \\
\hline 
CLEO & $\sim 10.6 $           & $\sigma({\rm B}\bar{\rm B}) \sim 1 \ {\rm nb}$ & $9.7 \times 10^6 {\rm B}\bar{\rm B} $\\
LEP  & $\sim 91$              & $\sigma(b\bar{b}) \sim 7  \ {\rm nb}$ & $0.9\times 10^6 b\bar{b}/$Expt\\
CDF/D0 & $1800$ & $\sigma(b\bar{b}) \sim 100 \ \mu{\rm b}$ & trigger dependent \\
\end{tabular}
\caption{\label{tab:nb} Numbers of $b$ hadrons accumlated at different experiments.}}

 The entire CLEO data sample consists of $9.7\times 10^6 \ \bbbar$ pairs
with $3.3\times 10^6 \ \bbbar$ pairs accumulated with the
CLEO II detector configuration~\cite{ref:cleoii}. The remaining
$6.4\times 10^6 \ \bbbar$ pairs were accumulated with the CLEO II.V
detector configuration~\cite{ref:cleoii.v}. The innermost tracking chamber
was replaced by a three-layer, double-sided silicon vertex detector
and the argon-ethane gas mixture in the large drift chamber was replaced
by helium-propane to convert CLEO II to CLEO II.V. The latter of these 
detector modifications improves the charged kaon and pion separation
both in terms of specific ionization $dE/dx$ measured in the drift chamber
and the momentum resolution.

\section{Charmless hadronic B decays}
 Charmless hadronic B decay candidates are selected by requiring
$|\EB - \EBEAM| \equiv |\Delta E| < 200$ or $300$ MeV  
(The $\Delta E$ resolution varies from
15 to 25 MeV depending on the final state),
$5200 < \MB < 5300 \ {\rm MeV}$ 
($\sigma(\MB) \approx 2.5 \ {\rm MeV}$) and 
$|\cos(\theta_{\rm sphericity})|$ $ < 0.8$ or $0.9$~\cite{ref:sphericity}.
The $|\Delta E|$ and $|\cos(\theta_{\rm sphericity})|$ differ slightly
for the different decay modes.
The latter requirement exploits the difference in the shape of $\bbbar$
and continuum events. In addition loose resonance mass $M_{\rm res}$
and particle identification cuts are used where applicable. Yields
are extracted from the sample of resulting B candidates from an
unbinned maximum likelihood (ML) fit~\cite{ref:fisher} to the largely independent
variables $\MB$, $\Delta E$, ${\cal F}$, $dE/dx$, 
$M_{\rm res}$ and $\cos\theta_{\rm helicity}$, 
where ${\cal F}$ is a Fisher discriminant combining 11 event shape 
variables~\cite{ref:fisher},
$M_{\rm res}$ is the $\rho,{\rm K}^*,\eta,\eta'$ or $\omega$ candidate
mass and $\cos\theta_{\rm helicity}$ is the helicity angle appropriate
for pseudoscalar $\to$ pseudoscalar,vector decays. In order to graphically
present the results, cuts are applied to all variables in the fit except 
for the variable being plotted. These cuts generally reduce the signal
efficiency by $\sim 50\%$ and the background by an order of magnitude.

\FIGURE[h]{\epsfig{figure=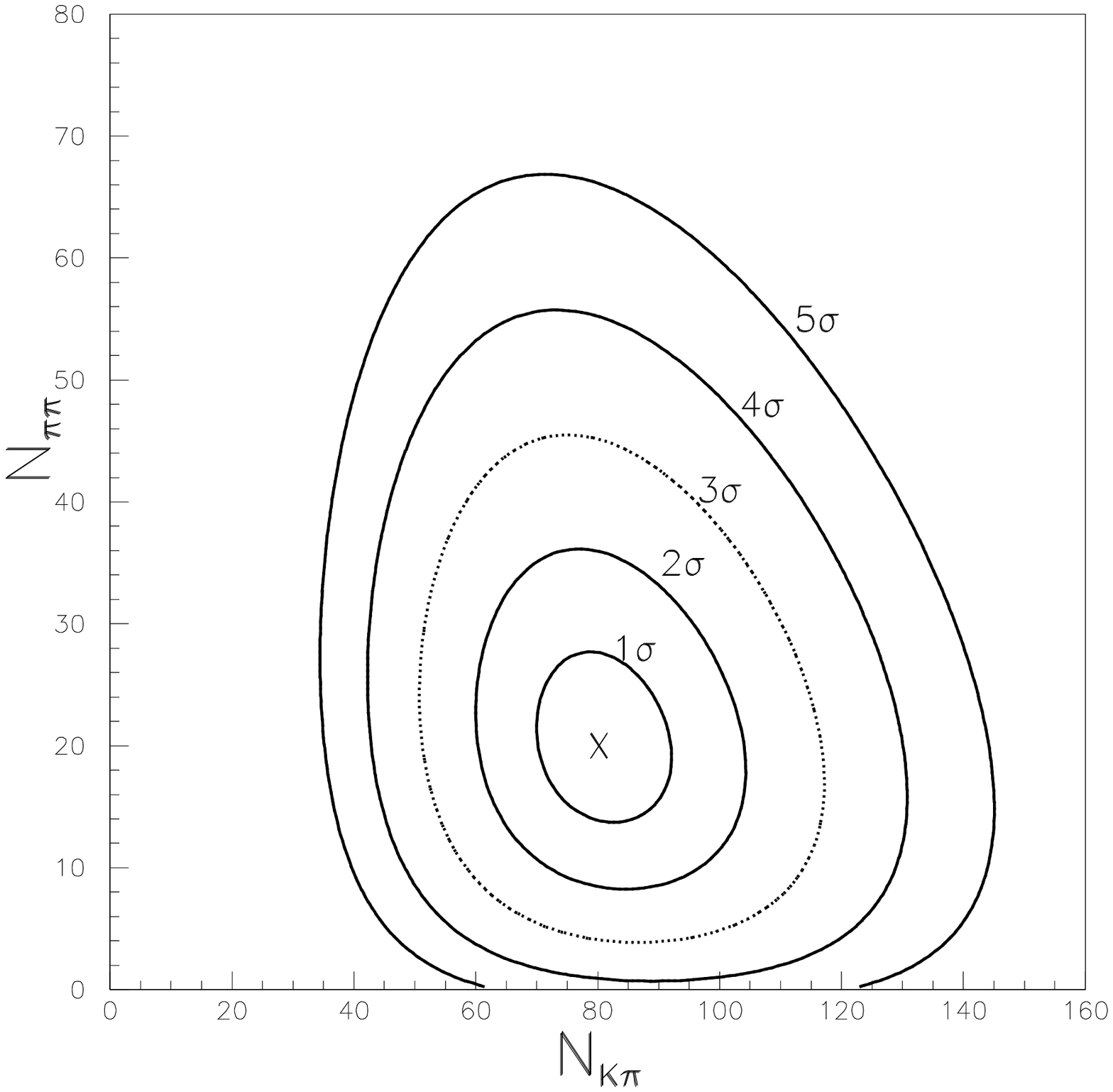,clip=,bbllx=30,bblly=120,bburx=530,bbury=650,width=0.7\linewidth}
\caption{\label{fig:kpi.contours}The likelihood contours in intervals of standard
deviations (statistical uncertainty only) for the 
$\BZ\to\PPPM$ {\it vs.} 
$\BZ\to{\rm K}^{\pm}\pi^\mp$ 
 yields.}}
\FIGURE[h]{\epsfig{figure=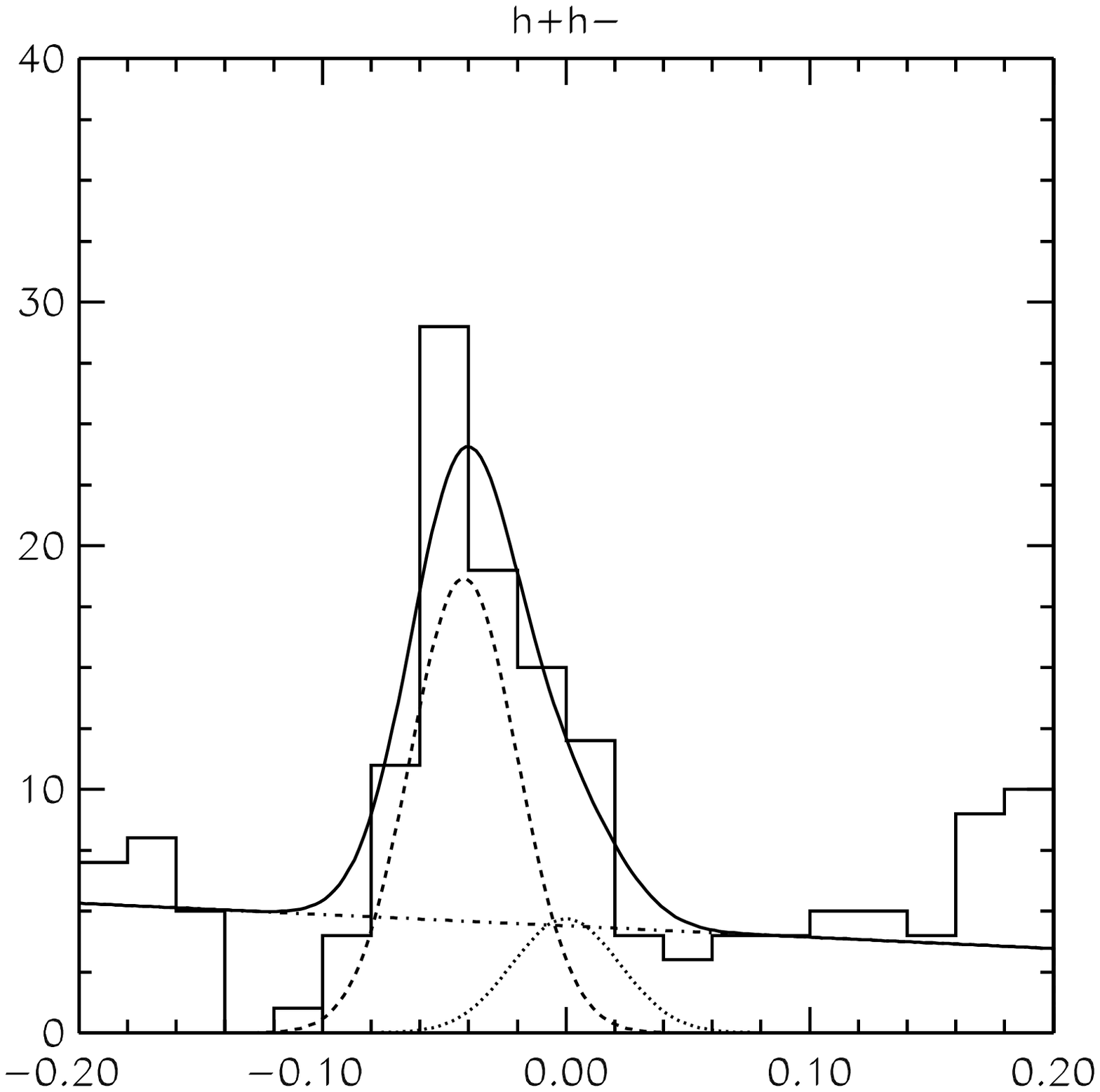,bbllx=000,bblly=100,bburx=700,bbury=625.550,width=0.907\linewidth}
\caption{\label{fig:kpi.deltae}The difference in the $\BZ\to h^+ h^-$ candidate energy
and $\EBEAM$ in GeV when both $\BZ$ daughter candidates are assigned the charged
pion mass. The solid line represents the full fit, the dashed line represents
the $\KPPM$ component, the dotted line represents the smaller $\PPPM$
component, the dot-dash line represents the background and
the histogram represents the data.
}}

 The preliminary results from the ML fit to $\BZ \to h^+\pi^-$
candidates in $9.7\times 10^6 \ \bbbar$ pairs are shown in
figures~\ref{fig:kpi.contours} and \ref{fig:kpi.deltae}. The measured
yields for ${\rm K}^+\pi^-$ and $\pi^+\pi^-$ are
$\YBZTOKPPM$ and $\YBZTOPPPM$ events, respectively.
This represents the first observation of the decay $\BZ\to\PPPM$.
Figure~\ref{fig:kpi.contours} shows the likelihood contours in increments
of standard deviations (statistical uncertainty only); the $\BZ\to\PPPM$
is significant at over four standard deviations. Figure~\ref{fig:kpi.deltae}
demonstrates one component of CLEO's ability to separate charged kaons
and pions; when a kaon is assigned the pion mass, the shift in
$\Delta E$ is $-42$ MeV ($\sigma(\Delta E) = 25[20] \ {\rm MeV}$
for CLEO II [CLEO II.V]). The ${\rm K}/\pi$ separation in $dE/dx$ at
$|\vec p_h| \approx 2.5 \ {\rm GeV}$ is 1.7[2.0]$\sigma$ in CLEO II [CLEO II.V].
The preliminary results for the branching fractions of $\BZ \to h^+h^-$ decays
are $\BRBZTOKPPM$, $\BRBZTOPPPM$ and $<\BRBZTOKPKM$ at  90\% CL for
$h^+h^- = \KPPM,\ \PPPM,\ {\rm and}\ \KPKM$, respectively~\cite{ref:kpi.lp99}~\footnote{All CLEO
results presented here assume 
$\Br(\Upsilon(4{\rm S}) \to \BZ\BZB) =\Br(\Upsilon(4{\rm S}) \to \BP\BM) = 0.5$}.

 The preliminary results for the related decays $\BP\to\KZ h^+$ and
$\BP\to h^+\PZ$ are 
$\Br(\BP\to\KZPP) = \BRBPTOKZPP$, 
$\Br(\BP\to\KZKP) < \BRBPTOKZKP$ at 90\% CL, 
$\Br(\BP\to\KPPZ) = \BRBPTOKPPZ$ and 
$\Br(\BP\to\PPPZ) < \BRBPTOPPPZ$ at 90\% CL~\cite{ref:kpi.lp99}. 
Significant signals are seen in
decays to 
$\KZPP$ and $\KPPZ$ but not in $\PPPZ$ or $\KZPP$. The relatively low
values of $\Br(\BZ\to\PPPM)$ and $\Br(\BP\to\PPPZ)$ with respect to
$\BZ\to\KPPM$ and $\BP\to\KPPZ$ indicate that there will be substantial
experimental difficulties in addition to the theoretical uncertainties
in measuring the angle $\alpha$ of the unitarity triangle 
(UT)~\cite{ref:pipi.alpha}.

\FIGURE[h]{\epsfig{figure=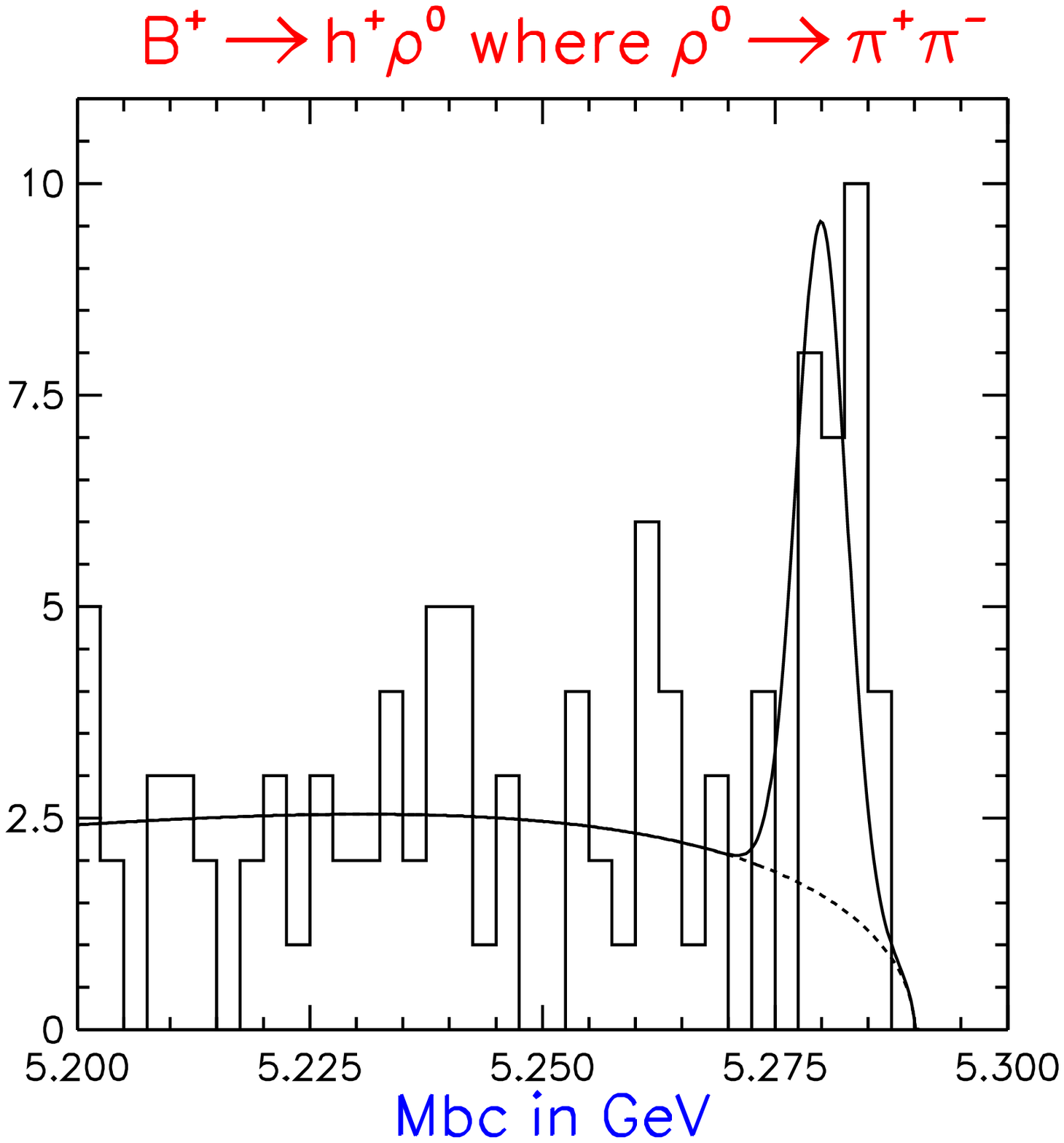,bbllx=100,bblly=100,bburx=650,bbury=623.575,width=0.907\linewidth}
\caption{\label{fig:rhopi.mb}The beam-constrained mass $\MB$ 
distribution for $\BP\to\RZPP$ candidates. The solid line represents the
full fit, the dashed line represents the background and the histogram
represents the data.
}}

\FIGURE[h]{\epsfig{figure=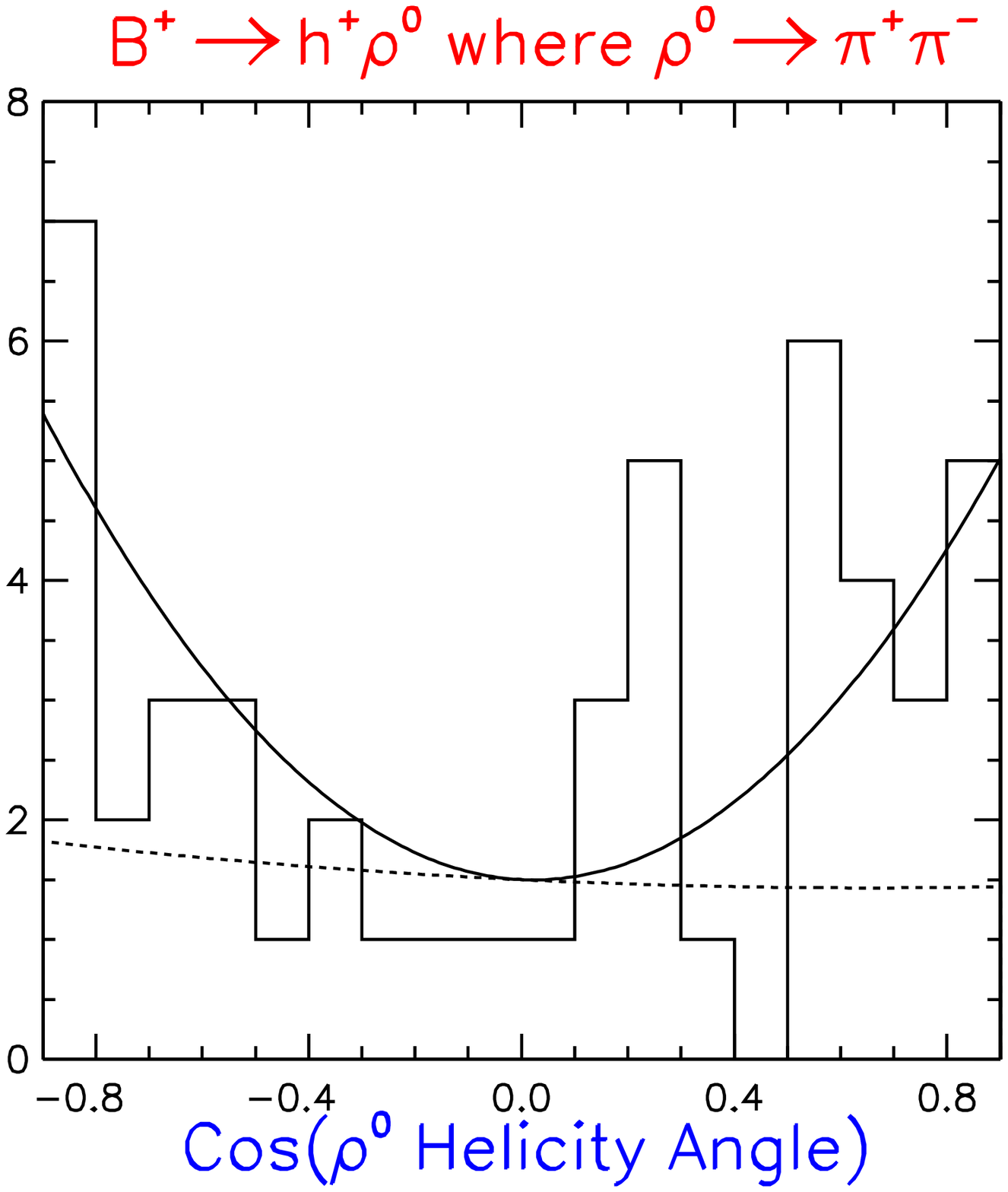,bbllx=100,bblly=100,bburx=650,bbury=623.575,width=0.907\linewidth}
\caption{\label{fig:rhopi.coshel}The $\cos\theta_{\rm helicity}$
distribution for $\BP\to\RZPP$ candidates.}}

 Figures~\ref{fig:rhopi.mb} and \ref{fig:rhopi.coshel} shows the beam-constrained
mass and $\cos\theta_{\rm helicity}$ distributions for
$\BP\to\rho^0 h^+$ candidates in $5.8\times 10^6 \ \bbbar$. Both
distributions show evidence of a $\BP$ signal. The yields
from the ML fit are
$\YBPTORZPP$ and $\YBPTORZKP$ for $\RZPP$ and $\RZKP$, respectively,
 corresponding to 
$\Br(\BP\to\RZPP) = \BRBPTORZPP$ and $\Br(\BP\to\RZKP) < \BRBPTORZKP$ at
90\% CL (preliminary).

\FIGURE[h]{\epsfig{figure=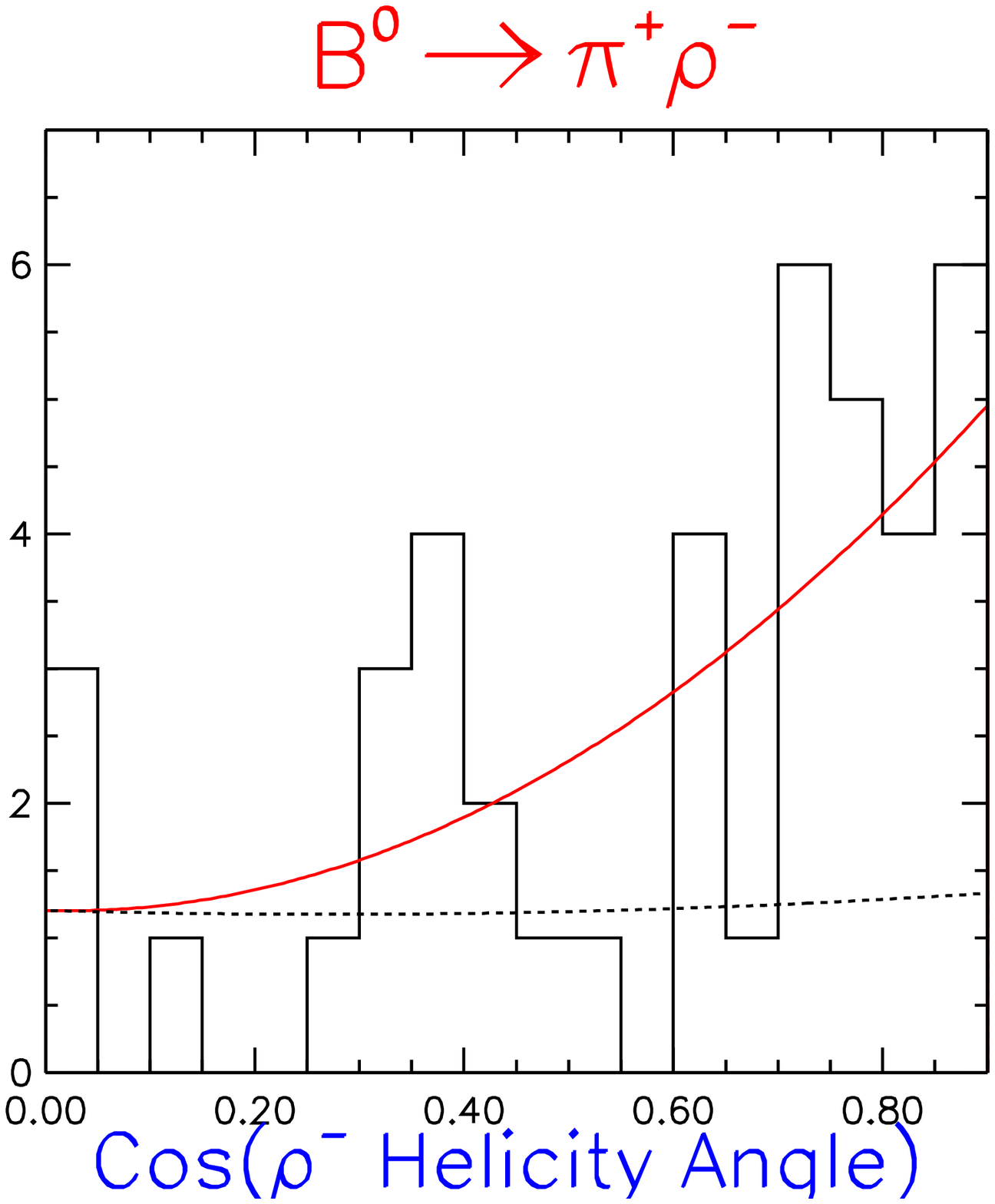,bbllx=100,bblly=100,bburx=650,bbury=623.575,width=0.907\linewidth}
\caption{\label{fig:rmpp.coshel}The $\cos\theta_{\rm helicity}$ distribution 
for $\BZ\to\RMPP$ candidates.}}

\FIGURE[h]{\epsfig{figure=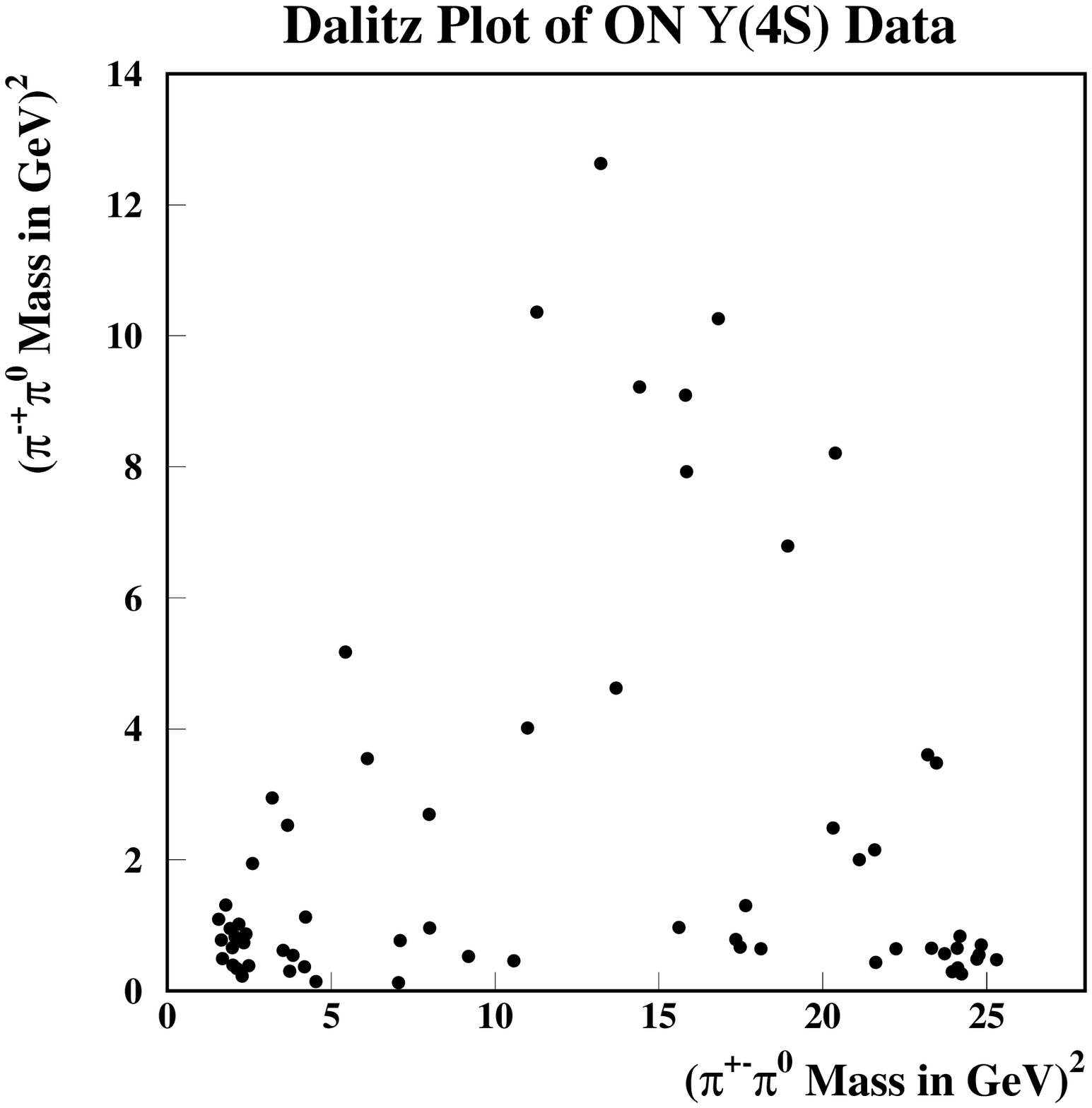,bbllx=0,bblly=0,bburx=550,bbury=550,width=0.907\linewidth}
\caption{\label{fig:rmpp.dalitz}The Dalitz distribution 
($M^2(\pi^\pm\pi^0)$ {\it vs.} $M^2(\pi^\mp\pi^0)$) for $\BZ\to\pi^+\pi^-\pi^0$
candidates. To improve the visibility of the $\rho$ resonance, the plot
is ``folded'': the larger (smaller) of $M^2(\pi\pi^0)$
is plotted on the horizontal (vertical) axis.}}

 For the related decay, $\BZ\to\rho^\pm h^\mp$, only positive values
of $\cos\theta_{\rm helicity}$  are considered 
(Fig.~\ref{fig:rmpp.coshel}). This serves to suppress backgrounds since it selects 
both a high momentum neutral pion that has less combinatorial background and
a low momentum charged pion that is well-separated from ${\rm K}^\pm$
by $dE/dx$ thus reducing potential backgrounds from $\BZ\to{\rm K}^{*+}\pi^-$.
The preliminary results are $\Br(\BZ\to\RMPP) = \BRBZTORMPP$
and $\Br(\BZ\to\RMKP) < \BRBZTORMKP$ at 90\% CL in $7.0\times 10^6\ \bbbar$.
In principle, the angle $\alpha$ could be measured to a precision of
$\sim 6^\circ$ with a likelihood fit to the proper time-dependence of
the $\pi^+\pi^-\pi^0$ Dalitz distribution with $\sim 1000$ 
{\em background-free} $\BZ\to\pi^+\pi^-\pi^0$ events~\cite{ref:snyder.quinn}.
Given CLEO's result, such a measurement would require over 100 fb$^{-1}$
at an asymmetric, ${\rm e}^+{\rm e}^-$ B-factory and would be complicated
by the backgrounds as shown in figure~\ref{fig:rmpp.dalitz}. In addition,
the region of the Dalitz distribution most sensitive to $\alpha$ due to
the interference between $\BZ\to\rho^+\pi^-$ and $\BZ\to\rho^-\pi^+$ occurs
when $M^2(\pi^+\PZ) \sim M^2(\pi^-\PZ) \sim M^2(\rho)$ where the backgrounds
to neutral pions are largest.

\FIGURE[h]{\epsfig{figure=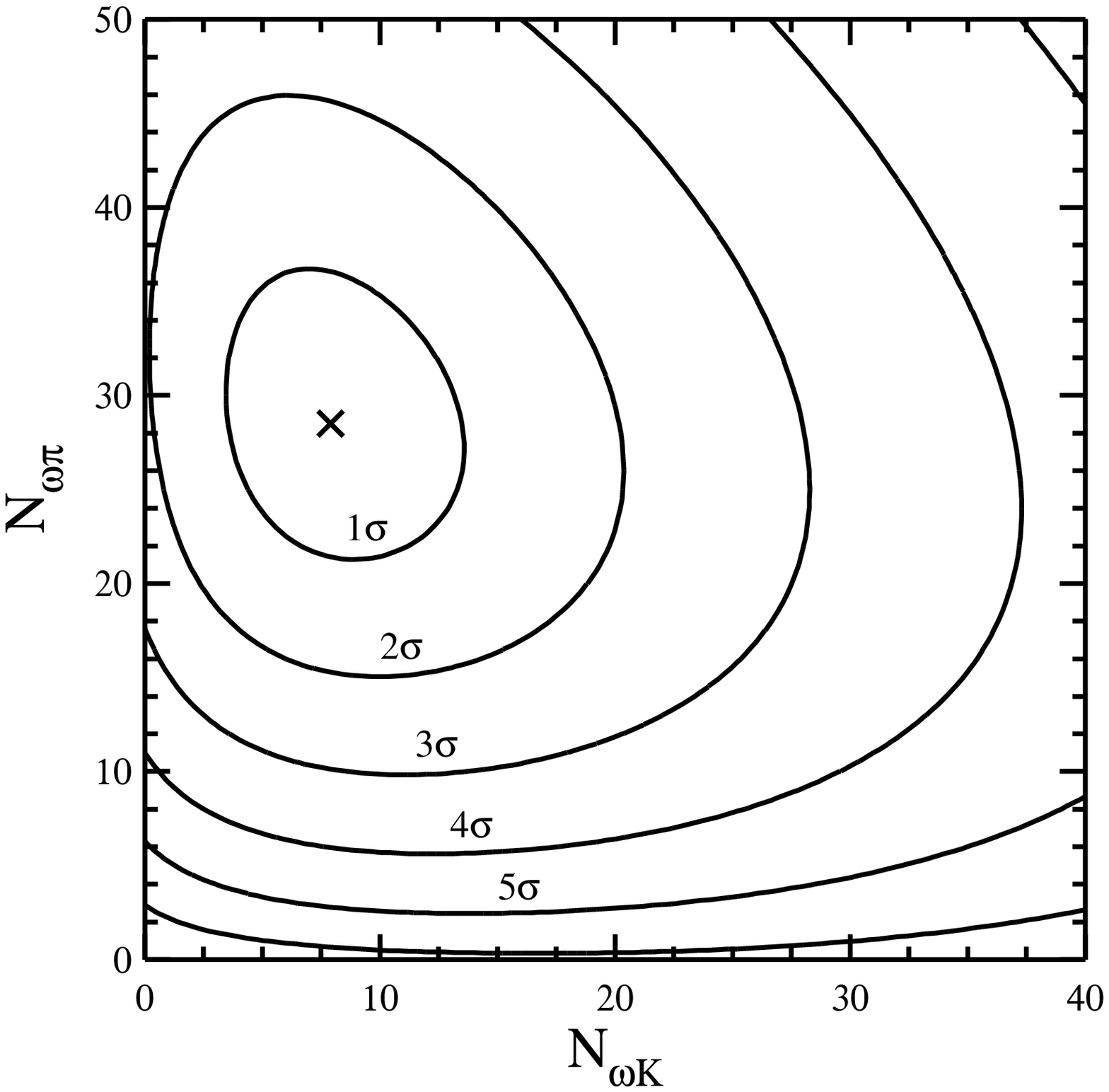,bbllx=000,bblly=100,bburx=650,bbury=580,width=0.907\linewidth}
\caption{\label{fig:omgpi.contours}The likelihood contours for the 
$\BP\to\omega\PP$ {\it vs.} $\BP\to\omega\KP$ yields.}}

 The likelihood contours for $\BP\to\omega h^+$ are shown in 
Figure~\ref{fig:omgpi.contours} for $9.7\times 10^6\ \bbbar$. The yield of
$\BP\to\OMGPP$ $(\OMGKP)$  is determined to be 
$\YBPTOOMGPP$ $(\YBPTOOMGKP)$ corresponding to
$\Br(\BP\to\OMGPP) = \BRBPTOOMGPP$ and 
$\Br(\BP\to\OMGKP) < \BRBPTOOMGKP$ at 90\% CL~\cite{ref:omegapi.LP99}. 
The $\BP\to\OMGPP$
branching fraction is consistent with that of $\BP\to\RZPP$ as expected
from isospin. The limit on the $\BP\to\OMGKP$ branching fraction is somewhat
at odds with CLEO's earlier reported value of 
$(1.5{}^{+0.7}_{-0.6}\pm 0.2)\times 10^{-5}$ in the $3.3\times 10^6\ \bbbar$
of the CLEO II data sample~\cite{ref:old.omegaK}. Two factors are responsible
for this change. The CLEO II data sample was re-analyzed with improved
calibration and track-fitting allowing an extension of the geometric
acceptance and track quality requirements resulting in an increase in reconstruction
efficiency of 10 to 20\%. The re-analysis reduced the $\OMGKP$ yield
and increased the $\OMGPP$ yield. In addition, there were very few
$\OMGKP$ candidates found in the CLEO II.V data sample ($6.4\times 10^6\ \bbbar$).

\FIGURE[h]{\epsfig{figure=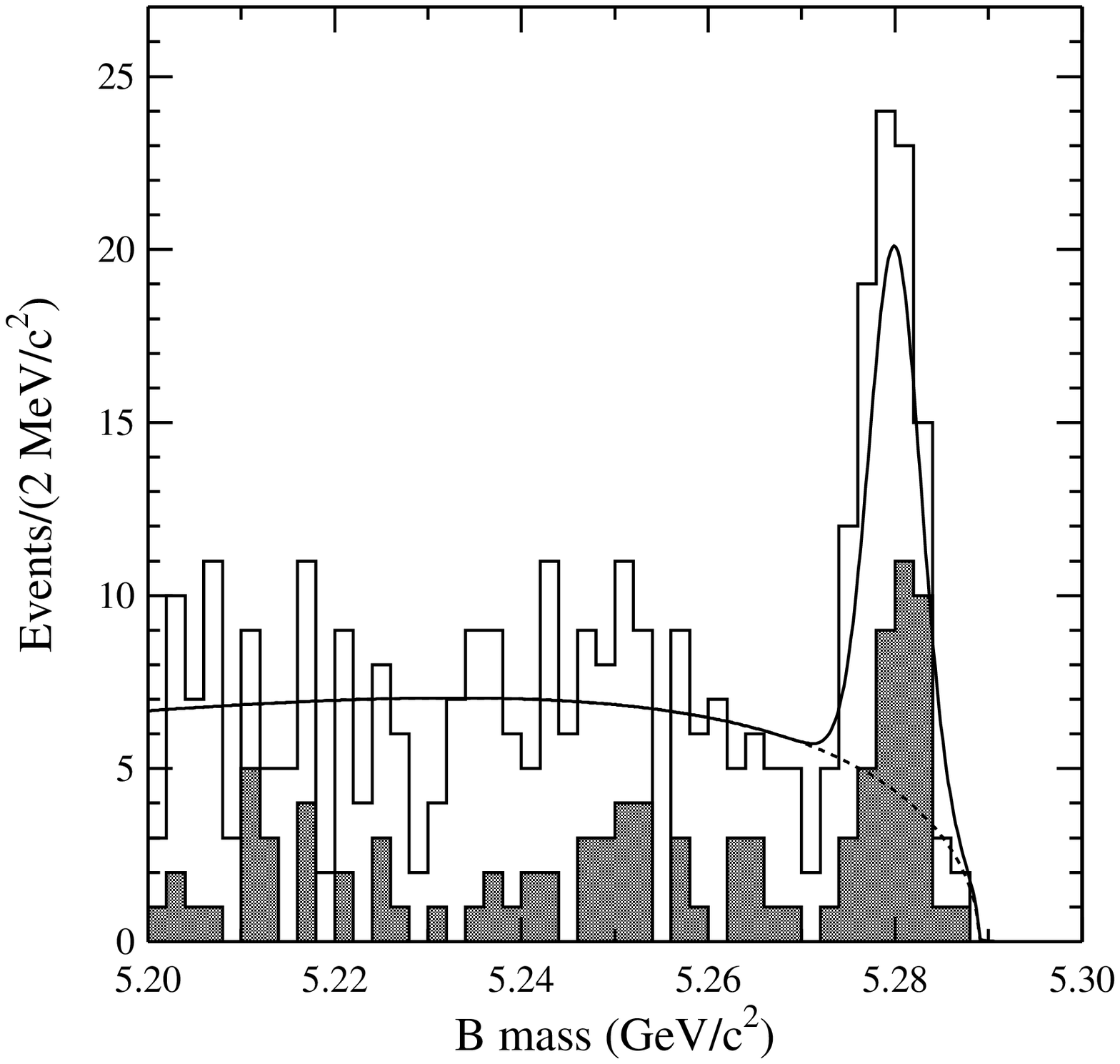,bbllx=040,bblly=120,bburx=540,bbury=620,width=0.907\linewidth}
\caption{\label{fig:etaprimeKP.mb}The beam-constrained mass for $\BP\to\eta'\KP$ candidates.
The shaded (white) region represents $\eta'$ candidates reconstructed in 
the $\eta'\to\PP\PM\eta$ ($\rho\gamma$) decay modes.
}}
\FIGURE[h]{\epsfig{figure=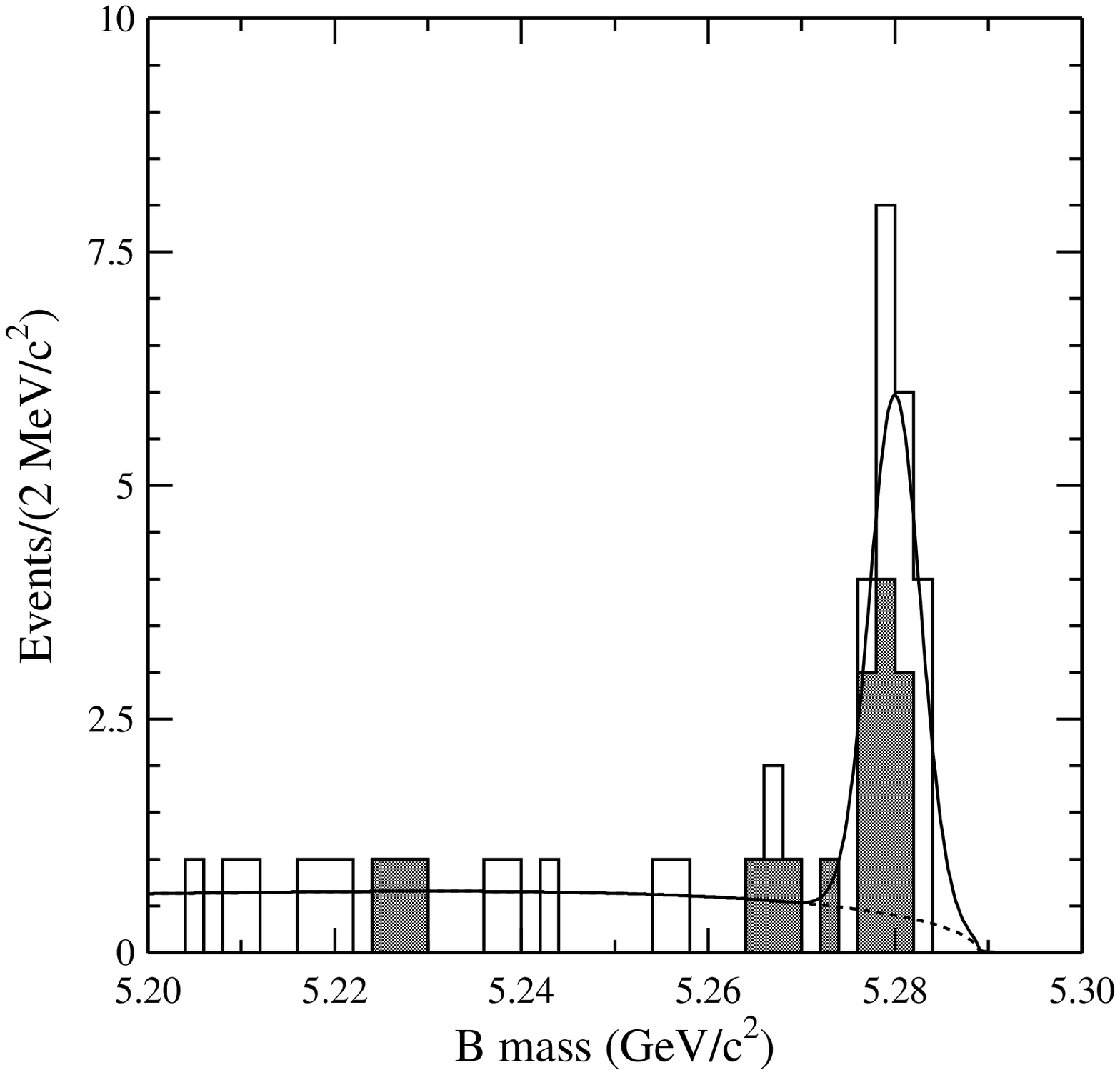,bbllx=40,bblly=120,bburx=540,bbury=620,width=0.907\linewidth}
\caption{\label{fig:etaprimeKZ.mb}The beam-constrained mass for $\BZ\to\eta'\KS$ candidates.
The shaded (white) region represents $\eta'$ candidates reconstructed in 
the $\eta'\to\PP\PM\eta$ ($\rho\gamma$) decay modes.
}}
  The beam-constrained mass distributions for $\BP\to\ETAPKP$ and 
$\BZ\to\ETAPKZ$ are shown in figures~\ref{fig:etaprimeKP.mb} and \ref{fig:etaprimeKZ.mb}.
The corresponding, preliminary branching fractions from $9.7\times 10^6\ \bbbar$
are $\Br(\BP\to\ETAPKP)=\BRBPTOETAPKP$, $\Br(\BZ\to\ETAPKZ)=\BRBZTOETAPKZ$
and $\Br(\BP\to\ETAPPP) < \BRBPTOETAPPP$ at 90\% CL~\cite{ref:etaK.LP99}. 
These surprisingly
large branching fractions are compared to predictions in table~\ref{tab:eta}.
Disparate explanations of the large ${\rm B}\to \eta'{\rm K}$ branching
fractions have been offered such as a large gluonic~\cite{ref:etaK.glue}
or intrinsic charm~\cite{ref:etaK.cc} content of the $\eta'$. Lipkin
extended isospin and flavor symmetry treatment of B decays to include
final state interactions and proposed the sum rule~\cite{ref:etaK.lipkin}
\begin{eqnarray*}
\Br(\BP\to\ETAPKP) &+& \Br(\BP\to\eta{\rm K}^+) = \\
\Br(\BP\to\KPPZ) &+& \Br(\BP\to\KZPP) 
\end{eqnarray*}
\noindent which can be evaluated with CLEO measurements and limits at
90\% CL listed in tables~\ref{tab:eta} and \ref{tab:kpirho} (units are $10^{-5}$)
\begin{eqnarray*}
(\UBRBPTOETAPKP) &+& (<\UBRBPTOETAKP)   \stackrel{?}{=} \\
(\UBRBPTOKPPZ) &+& (\UBRBPTOKZPP) \ \ \  .
\end{eqnarray*}
\noindent  The equality is violated by more
than three standard deviations. Recent work~\cite{ref:etaK.penguin}
suggests that the large $\eta'{\rm K}$ and $\eta{\rm K}^*$ rates
with respect to $\eta'{\rm K}^*$ and $\eta{\rm K}$ are due to constructive
interference of two comparable penguin amplitudes rather than mechanisms
specific to the $\eta'$.

\TABLE[hbt]{
\begin{tabular}{lcccc}
Reaction                 & $\Br$ or UL $(\times 10^{-5})$ & Prediction& $N(\bbbar) (\times 10^6)$& Reference\\
\hline
$\BP\ra\eta'{\rm K}^+$   & $\UBRBPTOETAPKP$           & 2.1-4.1    & 9.7 & \cite{ref:etaK.LP99}  \\
$\BZ\ra\eta'\KZ$         & $\UBRBZTOETAPKZ$  & 2.1-4.1    & 9.7 & \cite{ref:etaK.LP99}  \\
$\BP\ra\eta'{\rm K}^{*+}$& $<\UBRBPTOETAPKSTP$  & 0.03-0.04  & 6.7 & \cite{ref:etaK.DPF99} \\
$\BZ\ra\eta'{\rm K}^{*0}$& $<\UBRBZTOETAPKSTZ$  & 0.01-0.04    & 6.7 & \cite{ref:etaK.DPF99} \\
\hline
$\BP\ra\eta{\rm K}^+$   & $< \UBRBPTOETAKP$  & 0.2-0.4    & 9.7 &  \cite{ref:etaK.LP99} \\
$\BZ\ra\eta\KZ$         & $< \UBRBZTOETAKZ$  & 0.2-0.4    & 9.7 &   \cite{ref:etaK.LP99}\\
$\BP\ra\eta{\rm K}^{*+}$& $\UBRBPTOETAKSTP$   & 0.2-0.3  & 9.7 &   \cite{ref:etaK.LP99}\\
$\BZ\ra\eta{\rm K}^{*0}$& $\UBRBZTOETAKSTZ$   & 0.2-0.3    & 9.7 &   \cite{ref:etaK.LP99}\\
\end{tabular}
\caption{\label{tab:eta} Preliminary CLEO results for $B \ra \eta(') {\rm K}^*$ compared to
predictions from Ref.~\protect\cite{ref:etaK.akl}.}}

\TABLE[hbt]{
\begin{tabular}{lc|lcc}
Decay        & $\Br\  {\rm or}\ {\rm UL}\ (\times 10^{-5})$ & 
Decay        & $\Br\  {\rm or}\ {\rm UL}\ (\times 10^{-5})$ & \\
\hline
$\BZ\ra\KPPM$& $\UBRBZTOKPPM$      &$\BZ\ra\KSTPPM$ &$\UBRBZTOKSTPPM$&\\
             &                     & $\BZ\ra\RMKP$   & $< \UBRBZTORMKP$  & \\
$\BP\ra\KPPZ$& $\UBRBPTOKPPZ$      &$\BP\ra\RZKP$ & $< \UBRBPTORZKP$ & \\
             &                     & $\BP\ra\OMGKP$  & $< \UBRBPTOOMGKP$&        \\ 
$\BP\ra\KZPP$& $\UBRBPTOKZPP$ &                 &                   & \\
\hline
$\BZ\ra\PPPM$& $\UBRBZTOPPPM$ & $\BZ\ra\RMPP$ & $\UBRBZTORMPP$& \\
$\BP\ra\PPPZ$&$<\UBRBPTOPPPZ$          & $\BP\ra\RZPP$ & $\UBRBPTORZPP$& \\
             &                & $\BP\ra\OMGPP$& $\UBRBPTOOMGPP$& \\
\hline
$\BZ\ra\KPKM$&$<\UBRBZTOKPKM$ & $\BZ\ra\KSTPKM$ & $< \UBRBZTOKSTPKM$& \\
$\BP\ra\KZKP$&$<\UBRBPTOKZKP$         &    &   & \\
\end{tabular}
\caption{\label{tab:kpirho} Summary of charmless hadronic branching fractions ($\Br$) or
upper limits (UL) at 90\% CL.}}

 The summary of CLEO's charmless hadronic B decay measurements is shown
in table~\ref{tab:kpirho}. There is now clear evidence of hadronic
$b\to u$ transistions ($\BZ\to\PPPM$, ${\rm B}\to \rho/\omega\pi$). The
pattern of 
$\Br({\rm B}\to{\rm K}\pi) \approx \Br({\rm B}\to \rho/\omega\pi) > \Br({\rm B}\to\pi\pi)$
is an indication of the constructive [destructive] interference between
tree and penguin contributions for $\BZ\to\KPPM$ and ${\rm B}\to \rho/\omega\pi$
[$\BZ\to\PPPM$].

\section{Radiative and leptonic $b$ decays} 
 Both ALEPH~\cite{ref:aleph.bsg} and CLEO~\cite{ref:cleo.bsg} employ
similar techniques to measure the $b\to s\gamma$ branching fraction.
Photons from $\pi^0$s and $\eta$s are vetoed in the selection of the high
energy photon candidate. ALEPH suppresses the light quark ($ucsd$) background
by imposing a lifetime-based $b$-tag in the hemisphere opposite the photon
candidate. At CLEO, the suppression of the $ucsd$ background is 
achieved by a neural network that utilizes event shape information.
Both experiments then form an $X_s$ candidate from a combination of
tracks, $\KS$ and $\PZ$ candidates that, when combined with the high
energy photon, produces the ``best'' $b$ hadron candidate. The photon energy
spectra from the remaining candiates are shown in figures~\ref{fig:cleo.bsg} 
and \ref{fig:aleph.bsg}.
The subtraction of the remaining background is accomplished
quite differently by the two experiments. CLEO takes advantage of the data
accumulated below the $\bbbar$ threshold to subtract the dominant
$ucsd$ background. ALEPH adjusts their simulation of the background processes
based on $b\to s\gamma$-poor 
regions of distributions that discriminate
between signal and background. This procedure reduces ALEPH's
sensitivity to their simulation but increases the systematic uncertainty.
The measured $b\to s\gamma$ branching fractions of
$\CLEOBSGAMMA$ and $\ALEPHBSGAMMA$ from CLEO and ALEPH, respectively, 
are in good agreement
with the SM calculation at next-to-leading order of 
$(3.28\pm0.33)\times 10^{-4}$~\cite{ref:bsg.theory}.
The CLEO result is preliminary and based on $3.3\times 10^6\ \bbbar$.

\FIGURE[h]{\epsfig{figure=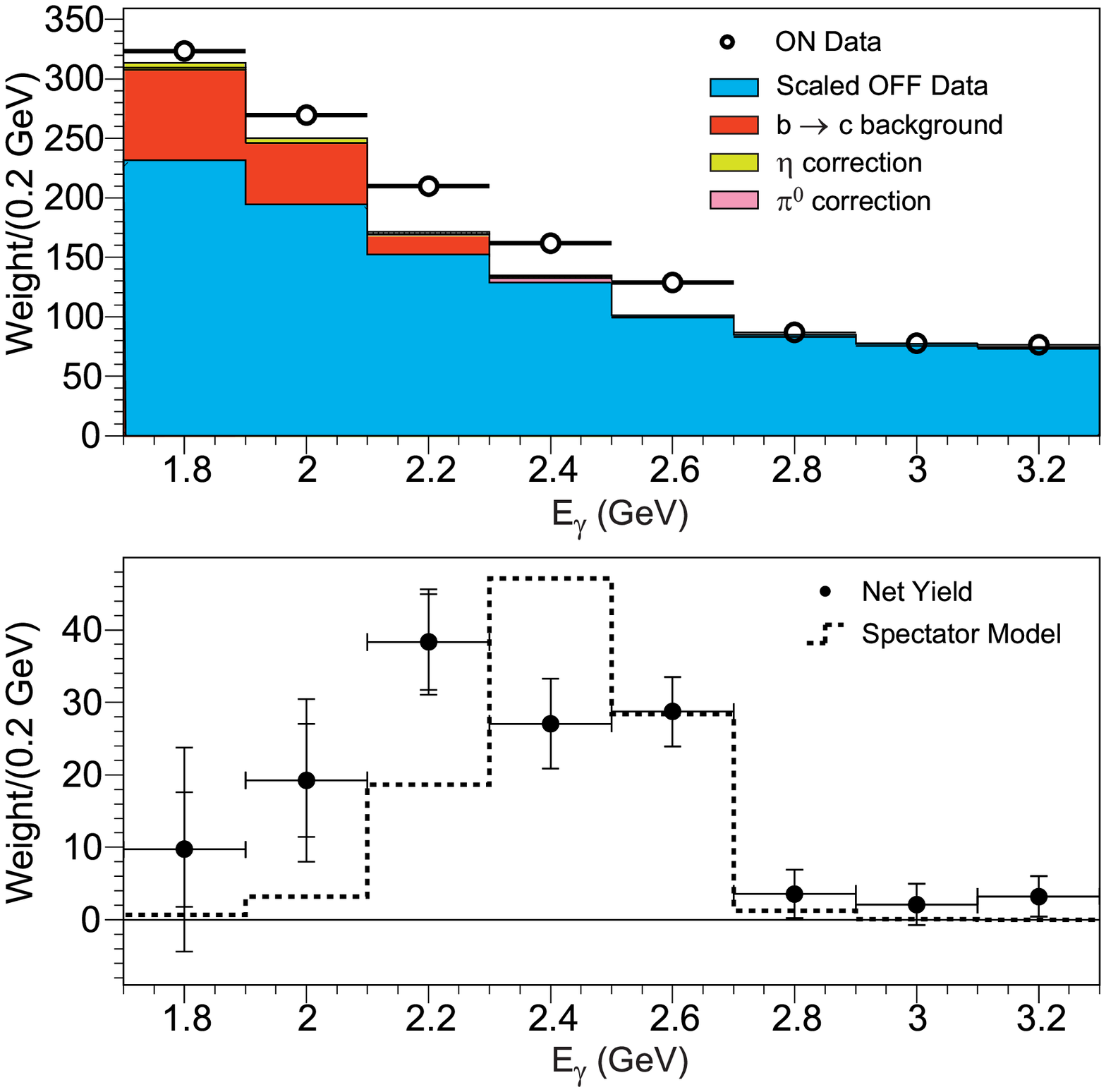,width=0.907\linewidth}
\caption{\label{fig:cleo.bsg}The photon energy spectrum for $b\to s\gamma$
candidates from the CLEO experiment. The upper (lower) figure
shows the distribution before (after) background subtraction.}}

\FIGURE[h]{\epsfig{figure=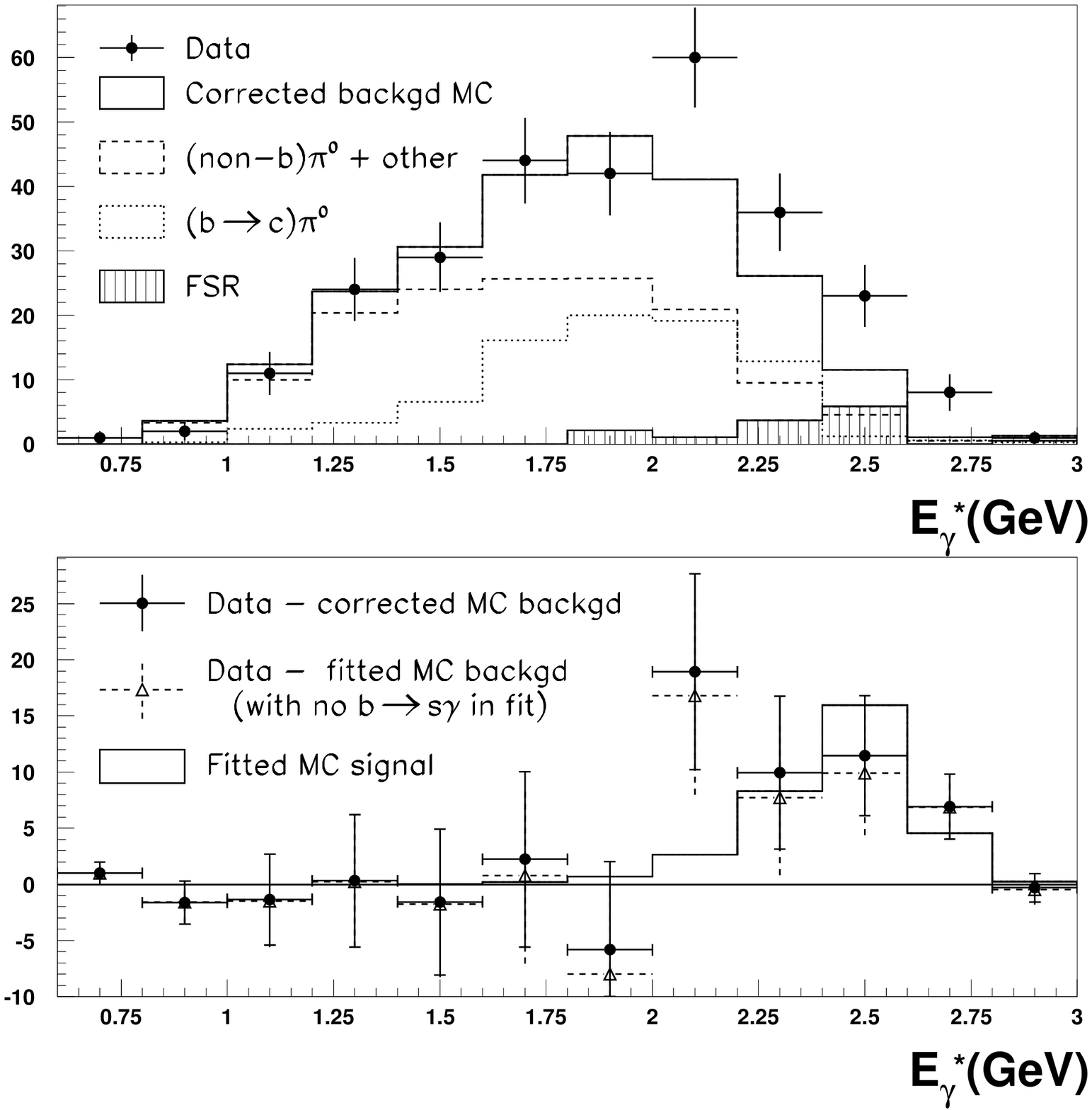,width=0.907\linewidth}
\caption{\label{fig:aleph.bsg}The photon energy spectrum for $b\to s\gamma$
candidates from the ALEPH experiment. The upper (lower) figure
shows the distribution before (after) background subtraction.}}

 Even though the $b\to s\gamma$ rate agrees with the SM calculation, many non-SM
effects could give rise to a sizeable ($\sim 40\%$) rate 
asymmetry~\cite{ref:bsg.asym}. CLEO extends their $X_s$ reconstruction method 
to ``tag'' the $b$-flavor of the final state of the $b\to s\gamma$ decay
to produce a measurement of the asymmetry
\begin{eqnarray}
{\cal A}^0 \equiv
     \frac {\Gamma(b\rightarrow s\gamma) - \Gamma(\bar{b}\rightarrow \bar{s}\gamma)}
           {\Gamma(b\rightarrow s\gamma) + \Gamma(\bar{b}\rightarrow \bar{s}\gamma)}
 \ \ \  .
\end{eqnarray}
\noindent Monte Carlo is used to determine the rate of the three
possible outcomes of tagging: 1) $b$ flavor determinable and 
correctly assigned, 2) $b$ flavor determinable and incorrectly
assigned or 3) $b$ flavor not determinable. The preliminary CLEO
result, based on $3.3\times 10^6\ \bbbar$, for the measured asymmetry,
after both additive and multiplicative corrections, is
${\cal A}^0 \approx 
 {\cal A}^{\rm meas} = 
 (0.16\pm 0.14_{\rm stat}\pm 0.05_{\rm syst})\times(1.000\pm 0.041)_{\rm syst}
$ or $-0.09 <  {\cal A}^{\rm meas} < 0.42 $ at 90\% CL.

 Hadron collider experiments are currently only sensitive to rare $b$
decays containing charged lepton pairs due to their triggering ability.
CDF exploits this sensitivity to search for the decays 
${\rm B}\to {\rm K}^{(*)}\mu^+\mu^-$~\cite{ref:cdf.kll}. 
The candidates for these decays
must form a common vertex ($\chi^2({\rm vertex}) < 20$) with a measured
decay length greater than 400 microns transverse to the $p\bar{p}$
collision axis. In addition the ${\rm B}\to {\rm K}^{(*)}\mu^+\mu^-$
candidates must satisfy an isolation requirement that takes advantage
of the hard fragmentation of $b$ quarks. Figures~\ref{fig:mll.v.mkpll}
and \ref{fig:mll.v.mksll} 
show the $M(\mu^+\mu^-)$ {\it vs.} $M({\rm K}^{(*)}\mu^+\mu^-)$
distributions for the $\BP\to {\rm K}^{+}\mu^+\mu^-$ and 
$\BZ\to {\rm K}^{*0}\mu^+\mu^-$ candidates. The candidates due to
${\rm B}\to \psi^{(')}{\rm K}^{(*)}$ decays dominate the distributions which
are otherwise relatively background free. The limits obtained by 
CDF~\cite{ref:cdf.kll} in table~\ref{tab:kll} 
are within an order of magnitude of the SM prediction 
and indicate that this decay should be observed with the increase
in luminosity ($\sim 2\ {\rm fb}^{-1}$) expected for the upcoming run.
The measurement of the dilepton mass spectrum and the lepton-pair
forward-backward asymmetry are important for the separation of the
long- and short-distance contributions to the decay~\cite{ref:kll.sd.ld}
and are sensitive to contributions from non-SM processes~\cite{ref:kll.np}, 
but will require another order of magnitude increase in luminosity.
\FIGURE[h]{\epsfig{figure=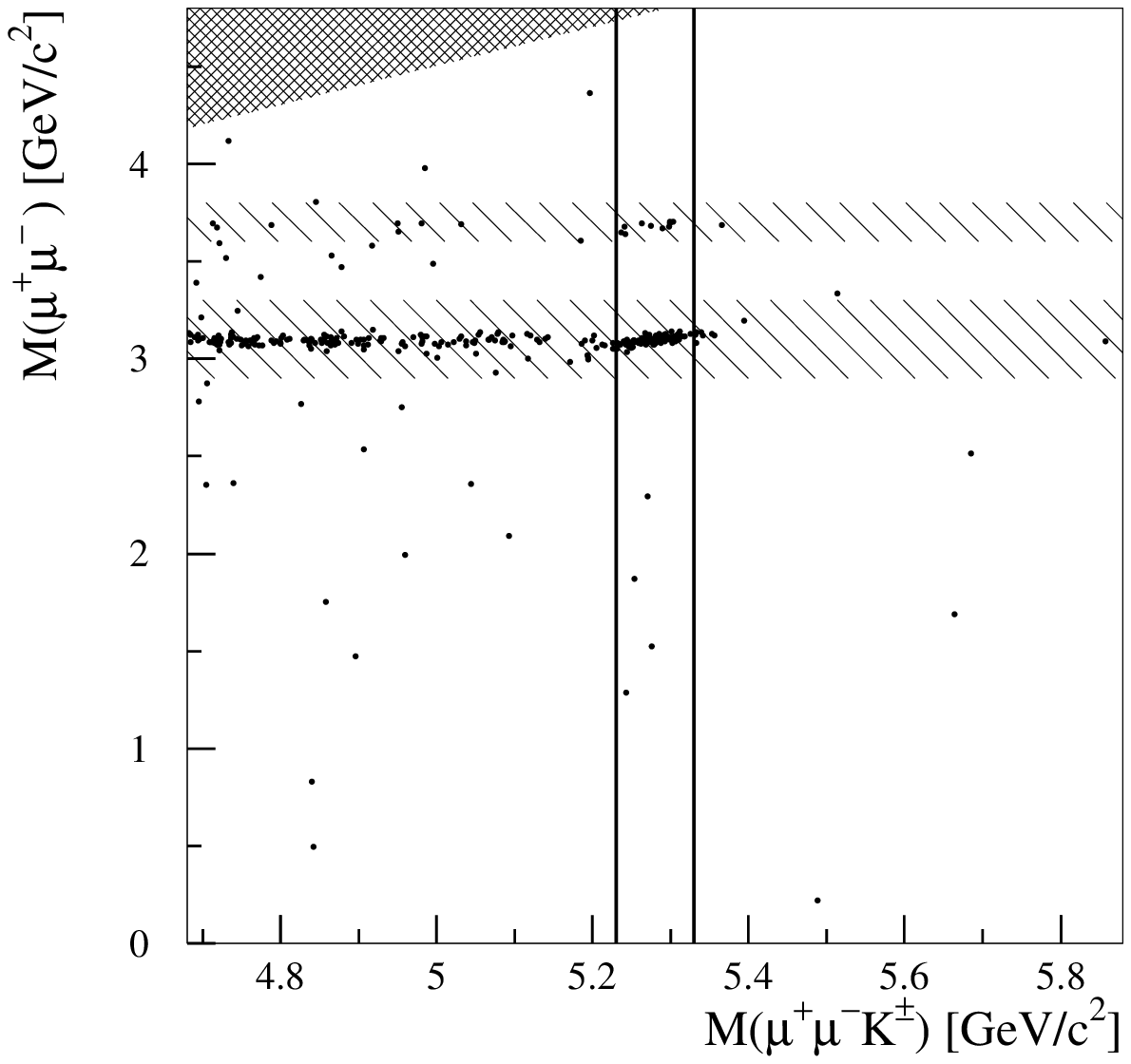,width=0.907\linewidth}
\caption{\label{fig:mll.v.mkpll} The $M(\mu^+\mu^-)$ {\it vs.} $M(\KP\mu^+\mu^-)$
distribution for $\BP\to\KP\mu^+\mu^-$ candidates from CDF.
The vertical lines delineate the signal region, the diagonal
hatching shows the excluded $\psi$ and $\psi'$ regions and
the shaded region is kinematically forbidden.
}}
\FIGURE[h]{\epsfig{figure=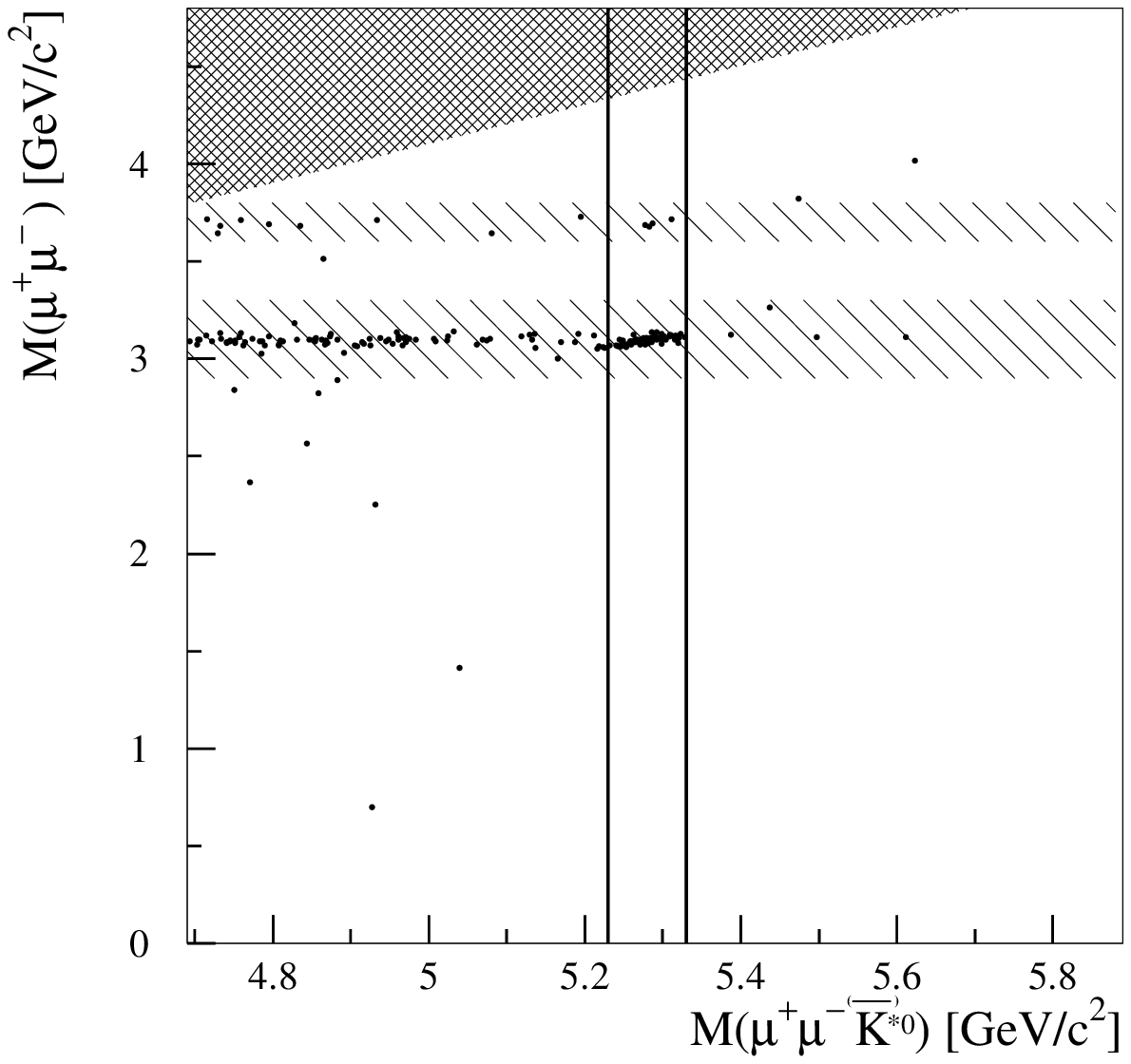,width=0.907\linewidth}
\caption{\label{fig:mll.v.mksll} The $M(\mu^+\mu^-)$ {\it vs.} $M({\rm K}^{*0}\mu^+\mu^-)$
distribution for $\BZ\to{\rm K}^{*0}\mu^+\mu^-$ candidates from CDF.}}

  Table~\ref{tab:kll} also lists results of searches for the inclusive
decay $b\to s\ell^+\ell^-$ by D0~\cite{ref:d0.sll} and CLEO~\cite{ref:cleo.sll}.
CLEO adapts their inclusive $X_s$ reconstruction from the
$b\to s\gamma$ measurement while D0 simply searches for high-mass lepton
pairs below the B mass and above the charmonium resonances. All measurements
are at least an order of magnitude higher than SM expectations.
\TABLE[hbpt]{
\begin{tabular}{lccc}
Expt               & $\Br(\BP\ra{\rm K}^+\ell^+\ell^-)$ & $\Br(\BZ\ra{\rm K}^{*0}\ell^+\ell^-)$ &$\Br(b \ra s\ell^+\ell^-)$\\
\hline
CDF $(\ell = \mu)$ & $ < 5.2 \times 10^{-6} $           & $< 4.0 \times 10^{-6}$                &\\
\hline
D0$(\ell = \mu)$   &                                    &                                       & $< 3.2 \times 10^{-4}$\\
\hline 
CLEO$(\ell = \mu)$ &  $ < 9.7 \times 10^{-6} $          & $< 9.5 \times 10^{-6}$                & $< 5.7 \times 10^{-5}$ \\
CLEO$(\ell = e)$   &  $ < 11 \times 10^{-6} $           & $< 13 \times 10^{-6}$                 & $< 5.8 \times 10^{-5}$ \\
CLEO$(\mu \ \& \ e)$&                                   &                                       &$< 4.2 \times 10^{-5}$  \\
\hline 
Std Model                 & $(0.3-0.7)\times 10^{-6}$          & $(1-4)\times 10^{-6}$& $\sim 6\times 10^{-6}$\\
\hline 
\hline 
CLEO\ $\mu^\pm e^\mp$ &                                 &                                       & $< 2.2 \times 10^{-5}$  \\
\hline 
\end{tabular}
\caption{\label{tab:kll} Summary of ${\rm B}\to {\rm K}^{(*)}\ell^+\ell^-$ 
and $b\to s\ell\ell$ searches.
All upper limits are at 90\% CL.}
}
\TABLE[hbpt]{
\begin{tabular}{lccc}
                           & CDF                     & CLEO                   & SM  \\
Decay                      & Upper limit at 95\% CL  & Upper limit at 90\% CL & expectation \\
\hline
${\rm B}^0_{\rm d} \ra \mu^+\mu^-$ & $<8.6\times 10^{-7}$ & $<5.9 \times 10^{-6}$ & $2\times 10^{-10}$\\
${\rm B}^0_{\rm d} \ra e^+e^-$     &                      & $<5.9 \times 10^{-6}$ & $3\times 10^{-15}$\\
${\rm B}^0_{\rm s} \ra \mu^+\mu^-$ & $<2.6\times 10^{-6}$ &                   & $4\times 10^{-9}$\\
\hline
${\rm B}^0_{\rm d} \ra e^\pm\mu^\mp$ & $<8.2\times 10^{-6}$&  $<5.9 \times 10^{-6}$& 0\\
${\rm B}^0_{\rm s} \ra e^\pm\mu^\mp$ & $<4.5\times 10^{-6}$&                   & 0\\
\hline 
\end{tabular}
\caption{\label{tab:ll} Summary of $\BZ\to\ell^+\ell^-$ searches.}
}

 Both CDF and CLEO have searched for the purely leptonic $b$ decays $\BZ\to\ell^+\ell^-$ 
as listed in table~\ref{tab:ll}. The current upper limits are at least
several orders of magnitude above the SM expectations. The SM 
${\rm B}_s^0\to\mu^+\mu^-$ decay rate should observable with the full expected 
luminosity from CDF's upcoming run if backgrounds can be reduced and the
overall detection efficiency enhanced by at least a factor of two.
\section{Conclusions}
 As the era of the $b$ factories begins, the current knowledge of
charmless hadronic $b$ decays indicates that independent measurement
of the angles $\alpha$ and $\gamma$ of the unitarity triangle will be
difficult. The relatively low rate of $\BZ\to\PPPM$ with respect to
$\BZ\to\KPPM$ will hamper efforts to measure the time-dependent $CP$
asymmetry of $\BZ\to\PPPM$. The apparently large penguin contributions
to ${\rm B}\to\pi\pi$ will also complicate the extraction of $\alpha$
from the measured asymmetry. An alternative proposal to measure $\alpha$
with ${\rm B}\to \pi\pi\pi$ decays will require years of running at the
asymmetric, ${\rm e}^+{\rm e}^-$ B-factories and may only be feasible
at $p\bar{p}$ colliders. The complications to the measurement of
$\alpha$ due to large penguin-tree interference may make it possible
to measure or bound the angle $\gamma$; however, current measurements
and methods are unable to provide meaningful bounds or are model dependent.
While $\sin 2\beta$ will almost surely be measured with significant
precision and reported at the 9$^{\rm th}$ International Symposium
on Heavy Flavor Physics, measurements of $\alpha$ and $\gamma$ of comparable
precision from rare $b$ decays may well have to wait for future conferences.


\end{document}